\documentclass[twocolumn]{aastex7}
\usepackage{graphicx}
\usepackage{epstopdf}
\usepackage{natbib}
\usepackage[hyphens]{url}

\newcommand{\ee}[1]{\mbox{${} \times 10^{#1}$}}
\newcommand{\eten}[1]{\mbox{$10^{#1}$}}

\newcommand{\am}{\mbox{\arcmin}}

\newcommand{\kms}{\mbox{km s$^{-1}$}}

\newcommand{\x}{\mbox{${}\times{}$}}

\newcommand{\sigmasfr}{\mbox{$\Sigma_{\rm SFR}$}}

\newcommand{\rgal}{\mbox{$R_{\rm Gal}$}}

\newcommand{\reff}{\mbox{$R_{\rm e}$}}  
\newcommand{\rbar}{\mbox{$R_{\rm bar}$}}  
\newcommand{\rdisk}{\mbox{$r_{\rm d}$}}  

\newcommand{\msun}{\mbox{M$_\odot$}}

\newcommand{\n}{\mbox{$n$}}

\newcommand{\mean}[1]{\mbox{$\langle#1\rangle$}} 

\newcommand{\alphaco}{\mbox{$\alpha_{\rm CO}$}} 

\newcommand{\ico}{\mbox{$I_{\rm CO}$}} 
\newcommand{\rtwoone}{\mbox{$R_{\rm 21}$}} 
 
\newcommand{\jj}[2]{\mbox{$J = #1\rightarrow#2$}}

\newcommand{\go}{\mbox{$G_0$}}

\newcommand{\mstar}{\mbox{$M_{\star}$}}

\newcommand{\msunyr}{\mbox{M$_\odot$ yr$^{-1}$}}

\newcommand{\spitzer}{\mbox{\rm{Spitzer}}}

\newcommand{\mw}{\mbox{\rm{Milky Way}}}

\shorttitle{Molecular Analogues for the Milky Way}
\shortauthors{Evans et al.}

\graphicspath{{./}{figures/}{../Analysis/}{../../MW\ Joins\ PHANGS/Analysis/}}

\begin{document}

\title{Molecular Gas Morphological Analogues for the Milky Way}

\correspondingauthor{Neal J. Evans II}
\email{nje@astro.as.utexas.edu}

\author[0000-0001-5175-1777]{Neal J. Evans II}
\affiliation{Department of Astronomy, The University of Texas at Austin,
2515 Speedway, Stop C1400, Austin, Texas 78712-1205, USA}
\email{nje@astro.as.utexas.edu}

\author[0000-0002-9120-5890]{Davide Elia}
\affiliation{INAF-IAPS, Via del Fosso del Cavaliere 100, I-00133 Roma, Italy}
\affiliation{INAF, Sezione di Lecce, Via per Arnesano, I-73100 Lecce, Italy}
\email{davide.elia@inaf.it}

\author[0000-0002-1423-2174]{Keith Hawkins}
\affiliation{Department of Astronomy, The University of Texas at Austin,
2515 Speedway, Stop C1400, Austin, Texas 78712-1205, USA}
\email{keithhawkins@utexas.edu}

\author[0000-0002-9333-387X]{Sophia Stuber}
\affiliation{National Astronomical Observatory of Japan, 2-21-1 Osawa, Mitaka, Tokyo 181-8588, Japan}
\affiliation{Max Planck Institute for Radio Astronomy, Auf dem Hügel 69, 53121 Bonn, Germany
}
\email{astro@sophiastuber.de}

\author[0000-0003-0378-4667]{Jiayi Sun}
\affiliation{Department of Physics and Astronomy, University of Kentucky, 506 Library Drive, Lexington, KY 40506, USA}
\email{Jiayi.Sun@uky.edu}

\begin{abstract}
 Complete catalogs of molecular clouds  in the Milky Way allow analysis of the molecular medium and the star formation properties of the Milky Way that closely follows the method used for nearby galaxies.  We explore whether the big dip in the radial distribution of molecular gas in the Milky Way is peculiar and find several other galaxies with similar patterns, all with similar morphological classifications \added{of YClxxGnR, indicating a clearly defined, long bar leading to a grand-design spiral.} 
This category is fairly rare among galaxies in the PHANGS sample, but all galaxies with this classification have some evidence for dips in the radial distribution of CO emission. The lengths of the bars correlate with the extents of the dips. The Milky Way and the other galaxies with dips have similar stellar masses and star formation rates, both lying near the high ends of the distributions for all PHANGS galaxies.

\end{abstract}


\defcitealias{2022AJ....164...43S}{S22}

\section{Introduction} \label{sec:intro}

Comparison of the \mw\ to nearby galaxies can be illuminating. While we have the advantage of spatial resolution for studies in the \mw, the edge-on view within the Galaxy makes it difficult to be sure of the large-scale properties, which are readily apparent in other galaxies. 
In addition, studies of molecular gas in our Galaxy are complementary to studies of its stellar distribution.  While unaffected by extinction, CO surveys rely on distances that are poorly constrained from rotation curves
\citep[e.g.,][]{2017A&A...601L...5R}. Studies of stars in the Milky Way, while providing a clearer picture of its spatial, chemical and dynamical nature, are hampered in the inner \mw\ by extinction, line-of-sight confusion, and the existence of multiple stellar populations
\citep[e.g.,][]{2016ARA&A..54..529B}. 

The time is right to place the stellar and molecular \mw\ in the extra-galactic context. 
Over the last decade, we have entered a data-rich era that has enabled us to more completely map the molecular and stellar components of the Milky Way 
\citep[e.g.,][]{MD17,GaiaDr3}.
Further surveys of its star formation and molecular gas properties are underway
\citep{2019ApJS..240....9S, 2020ApJS..246....7S, 2025arXiv250814547Z,2021MNRAS.500.3027D, 2022MNRAS.513..296D, 2022A&A...664A..84N}. In addition, the properties of the stellar component of the \mw\ are also being studied intensively. The recent explosion in data from large-scale astrometric \citep[Gaia,][]{GaiaDr3}, spectroscopic \citep[e.g., SDSS-V,][]{sdssv2025}, and photometric \citep[e.g., Gaia, and many others,][]{GaiaDr3} surveys, have ushered in a new data-rich industrial revolution transforming our broad understanding of the Milky Way from its stellar populations. 

These large-scale datasets have been used to uncover multiple components that make up the Galactic disk, which includes a ``thin" disk that is chemically, spatially, and dynamically distinct from a ``thick" disk \citep[e.g.,][and references therein]{Gilmore1983, Edvardsson1993, Feltzing2013, Hayden2015, Hawkins2015, Hayden2017, Imig2023}. Stellar population analysis in the Galactic thin disk has revealed a complex chemical, dynamical, and spatially distinct set of structures that build the underlying thin disk. Some of these complexities include spiral arms in both gas and young stellar populations \citep[e.g.,][]{Poggio2021, Poggio2022, Hawkins2023, Hackshaw2024, Jurado2026}, the existence of negative radial (and vertical) metallicity gradients across a wide range in ages \citep[e.g.,][]{Mayor1976,Andrievsky2002, Daflon2004, Bergemann2014,Xiang2015,Anders2017,Yan2019,Hawkins2023}, which implies inside-out formation \cite[e.g.,][]{Larson1976,Frankel2019}, and dynamically distinct substructures \citep[e.g.,][]{Hunt2018,Jurado2026}. While significant progress has been made in understanding the (chemical, spatial, and dynamical) structure of the Milky Way through the detailed analysis of large samples (10$^{5-7}$) its stars, a more complete picture of our Galaxy, in the context of others, will require a joint analysis of its stars and molecular gaseous components. 

Comparison of CO emission in the \mw\ to that in other galaxies may allow insight into the molecular morphology of the \mw.
The detailed images of molecular gas in the PHANGS sample 
\citep{2021ApJS..257...43L} 
have been used to construct molecular morphological classifications 
\citep{2023A&A...676A.113S}, hereafter referred to as Stuber classes.
This paper compares the radial distribution of molecular gas to those of galaxies in the PHANGS sample.

\section{Data Sets}\label{sec:data}

\subsection{Extragalactic Data}\label{sec:exgaldata}

The PHANGS-ALMA survey
\citep{2021ApJS..257...43L}  
has provided data on molecular clouds and star formation on $\sim 100$ pc scales for 90  galaxies.
The sample targeted nearby ($d < 20$ Mpc), low to moderately inclined ($i \lesssim 75\arcdeg$), relatively massive 
($\mstar/\msun) \gtrsim \eten{9.75}$) star forming (${\rm SFR}/{\mstar} > \eten{-11}$ yr$^{-1}$) galaxies, where \mstar\ is the stellar mass and SFR is the star formation rate.
Furthermore, a set of high-level measurements, including the large-scale distribution of gas mass and star formation rate across all PHANGS galaxies, was derived by
\citet[][hereafter \citetalias{2022AJ....164...43S}]{2022AJ....164...43S}.
Here we focus on the measured radial distribution of the annular averaged intensity of CO \jj21\ emission. 
We extract these measurements from the data tables\footnote{
The data repository is 
\url{https://www.canfar.net/storage/vault/list/phangs/RELEASES/Sun_etal_2022}
and we used the radial profile tables as part of the v4.0 public release.
} published by \citetalias{2022AJ....164...43S}.

\subsection{Milky Way data}\label{sec:mwdata}

The data on molecular gas in the \mw, for comparison to PHANGS, comes from the catalog of clouds defined by CO \jj10\ emission
\citep{MD17}.
\citet{2025ApJ...980..216E,2025ApJ...992..161E}\defcitealias{2025ApJ...980..216E}{E25} (hereafter \citetalias{2025ApJ...980..216E})
adjusted this catalog to use the same rotation curve as was used by the Hi-GAL survey and adjusted masses for a value of \alphaco\ that varied with metallicity, $Z$
\citep{2020ApJ...903..142G,2022ApJ...929L..18E}.
We constructed the average intensity of CO \jj10\ emission $\mean{\ico}$ by summing all the products of \ico\ and cloud area from the updated catalog in 
\citet{2025ApJ...992..161E} 
within an annular increment of Galactocentric radius (\rgal)
and then dividing by the area of the annulus. The annular width was 0.5~kpc and points are plotted at the midpoints to match the data in \citetalias{2022AJ....164...43S}.

\section{Is the Milky Way Odd?}\label{sec:bigdip}

The \mw\ has been described as a typical galaxy (``magnificently mediocre") based on its stellar mass and luminosity
\citep{2025arXiv250706989K}
and as an unusual ``green valley" galaxy
 \citep{2016ARA&A..54..529B}
based on its specific star formation rate. It falls squarely on the whole-galaxy Kennicutt-Schmidt relation between star formation rate and total gas surface densities
\citep{2012ARA&A..50..531K}. Milky Way analogues have been selected on the basis of many different properties
\citep[e.g.,][]{2001ASPC..231....2K,2019MNRAS.489.5030F,2019ApJ...872..106K,2024KosNT..30d..81V,2020MNRAS.491.3672B,2023MNRAS.521.5810Z,2025arXiv251118612K},
but comparisons based on the morphology of the molecular gas are new.

There is a widely shown {\it model} of the \mw\ based on \spitzer\ observations by the GLIMPSE team
\citep{2005ApJ...630L.149B}, showing a bar and well-defined spiral arms, but the structure of the inner \mw\ is still quite uncertain. For example,
\citet{2025AJ....170..355B}
found no evidence for spiral arms in the HI4PI data, and they suggest that the \mw\ is a flocculent galaxy.

The \mw\ has a dramatic feature in plots of molecular gas or SFR versus \rgal: a big dip from the CMZ that extends out to about 4~kpc, followed by a rise to a plateau extending to 5.5 kpc, then a steady decline to 20 kpc
\citep{2022ApJ...941..162E,2025ApJ...980..216E}.\footnote{The small local peak around 8 kpc is presumably caused by small local clouds that may be missed at larger distances.} Extrapolation of the trend outside 5.5 kpc back to the center is roughly consistent with the molecular density there. The value of \mean{\ico} in the dip is about an order of magnitude below the trend interpolated between the center and the outer regions.
The region of the dip coincides with the long bar identified by
\citet{2005ApJ...630L.149B}.
Estimates of the bar half-length range from 3.5 kpc 
\citep{2021A&A...656A.156Q, 2023MNRAS.520.4779L} 
to $5.0\pm0.2$ kpc
\citep{2015MNRAS.450.4050W, 2016ARA&A..54..529B}. 
\citet{2023MNRAS.520.4779L} emphasize that their bar length of 3.5 kpc is based on dynamical considerations and that a spiral arm starting at the end of the bar extending to 4.8 kpc could reconcile the different measurements.

It is not clear whether the relative absence of molecular gas from 1-4 kpc is caused by the bar excluding it or by incorrect distance assignment because models of galactic rotation have not properly accounted for the bar.
While a lack of star formation can be seen in two-dimensional images of the \mw\ \citep{2022ApJ...941..162E}, the gaps are more prominent on the near side of the center. 
The issues of distance determination are less for one-dimensional radial distributions, which are less affected by the kinematic distance ambiguity, so we concentrate on those.

\begin{figure}[h!]
\includegraphics[width=0.47\textwidth]{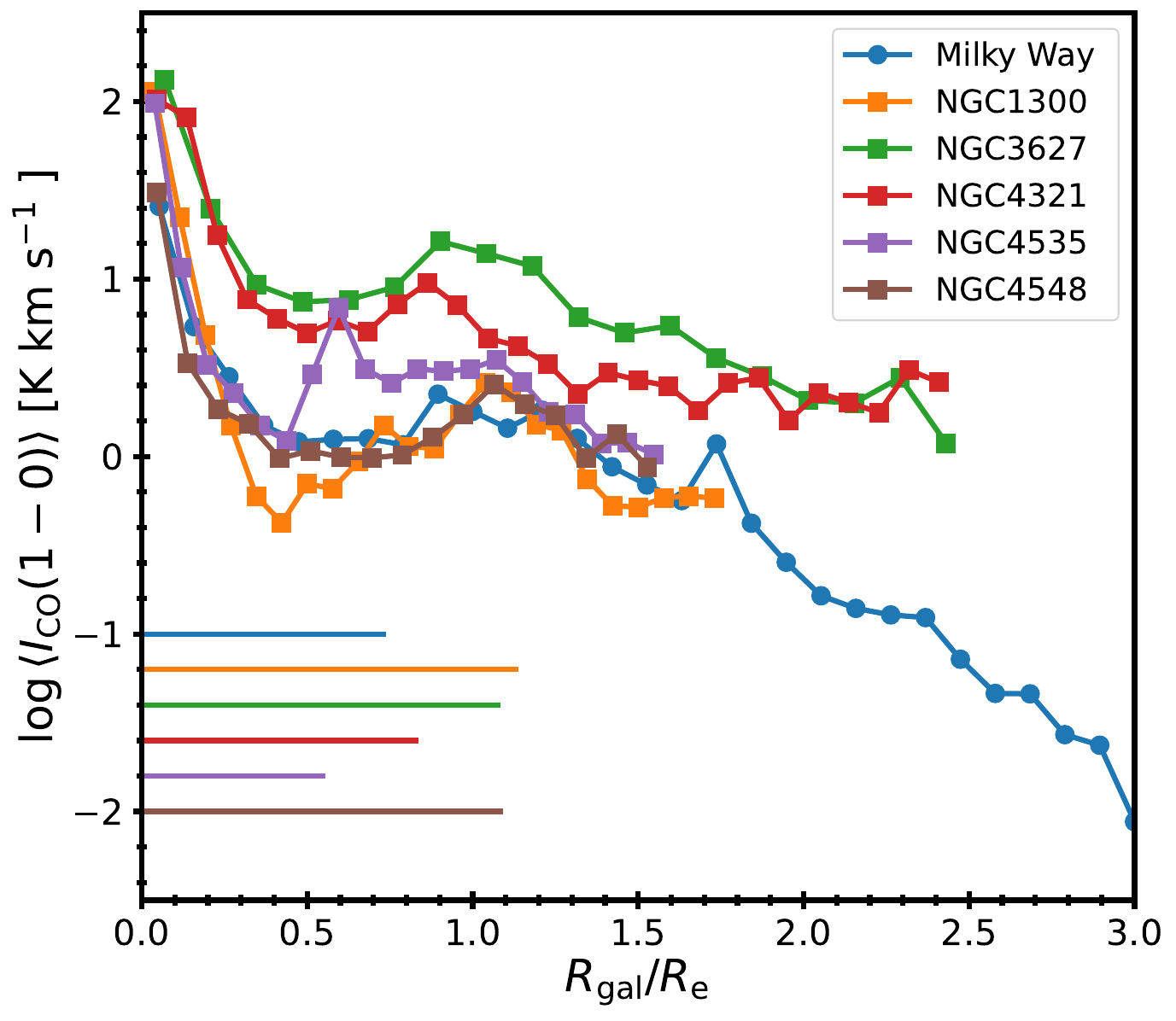}
\caption{The average CO \jj10\ intensity in radial bins is plotted for the \mw\  and for five PHANGS galaxies, converted from CO~\jj21\ \citepalias{2022AJ....164...43S} using a variable $R_{\rm 21}$ \citep{2025arXiv251005214S}. The radii are normalized to the effective radius of each galaxy, and the horizontal lines at the bottom left show the normalized bar length with the same color coding.}
\label{bigdipgals}
\end{figure}

Because bars and molecular gas distributions are much more easily discernible in  other galaxies, the PHANGS sample provides a good comparison set.
Big dips are not easily apparent in the plots where all galaxies are plotted. However, examination of all the galaxies in \citetalias{2022AJ....164...43S} with CO radial distributions out to 8 kpc (26 galaxies) found 5 other galaxies with visually similar big dips. They all have similar Stuber classifications \citep{2023A&A...676A.113S}: YClxxGnR, meaning clearly visible disks (Y) having clear (C), long (l) bars leading to grand-design spirals (G) without non-central rings (nR). The detailed bar shapes and presence or absence of a central ring vary among the five and are thus indicated as xx here. These are plotted, along with  the \mw\ in Figure \ref{bigdipgals}. To normalize for galaxy size and concentrate on shape, the plots are versus \rgal/\reff, \added{where \reff\ is the effective (half-light) radius}; plots versus \rgal\ alone \added{still indicate the similarity}. To place the PHANGS galaxies in context, the intensity in the CO \jj21\ line has been converted to the intensity in the \jj10\ line using a factor \rtwoone\ that depends on \rgal\ via the dependence on \sigmasfr\ \citep{2025arXiv251005214S}. 

\begin{figure}[h!]
\includegraphics[width=0.47\textwidth]{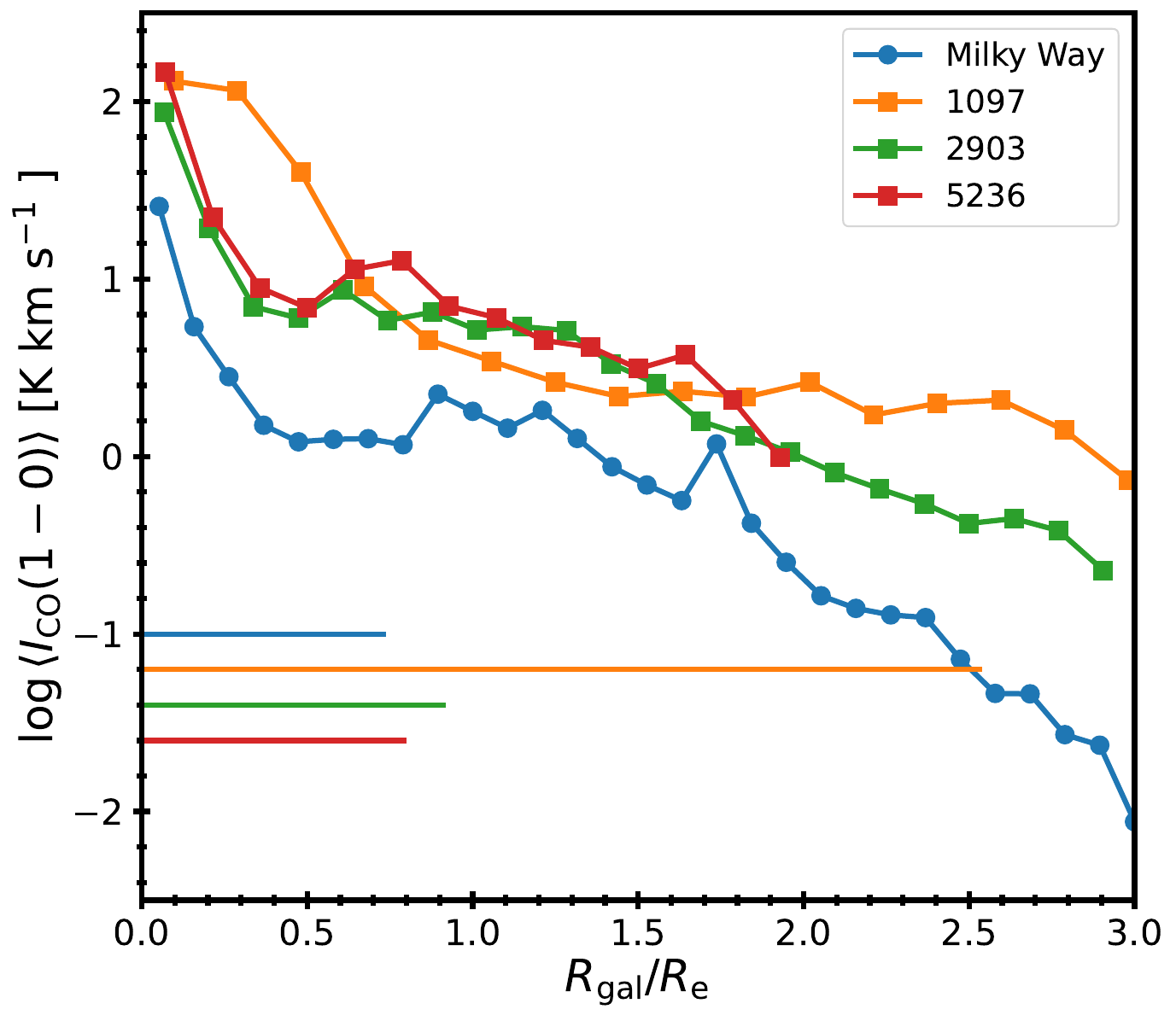}
\caption{The average CO \jj10\ intensity in radial bins is plotted for the \mw\  and for three PHANGS galaxies with similar classifications but less similar dips. The normalizations are the same as in Figure \ref{bigdipgals}.
}
\label{fig:smalldips}
\end{figure}

There are three other galaxies in the archive with the class of YClxxGnR. As plotted in Figure \ref{fig:smalldips}, they also show dips, though shallower. They were less apparent because the dips are shallow and either shorter or longer in $\rgal/\reff$  than the ``big dip" galaxies in Figure \ref{bigdipgals}. For both figures, the length of the bar for each galaxy is also plotted with the same color; the correlation of the dip extent with the length of the bar is apparent.

\begin{deluxetable*}{l c c c c c c c c c } 
\tablecaption{Properties of Galaxies Like the Milky Way \label{tab:bigdips}} 
\tablewidth{0pt} 
\tablehead{   
\colhead{Galaxy } &  \colhead{Distance}   & \colhead{$\log_{10} M_{*}$}   &  \colhead{$R_{\rm e}$}  &  \colhead{$R_{\rm bar}$} & \colhead{SFR} &  \colhead{$\log_{10}$ L(CO)} & Type & CVRHS  \\  
 & (Mpc) & (\msun) & (kpc) & (kpc) & (\msunyr) & (K \kms pc$^2$)  &     &   
  } 
\startdata 
\hline 
NGC 1300 & 19.00 & 10.61 & 6.50 & 7.40 & 1.20 & 8.73 & YCl1.5rGnR & (R\am)SB(s,bl,nrl)b G \cr 
NGC 3627 & 11.30 & 10.83 & 3.60 & 3.90 & 3.80 & 9.01 & YCl2.0nrGnR & SB\_x\_(s)b pec G \cr 
NGC 4321 & 15.20 & 10.75 & 5.50 & 4.60 & 3.60 & 9.14 & YCl3.5rGnR & SAB(rs,nr,nb)bc G \cr 
NGC 4535 & 15.80 & 10.53 & 6.30 & 3.50 & 2.20 & 8.90 & YCl2.0N/AGnR & SAB(s)c M \cr 
NGC 4548 & 16.20 & 10.69 & 5.40 & 5.90 & 0.52 & 8.51 & YCl1.0nrGnR & SB(rs,bl)\underline{a}b G \cr 
MW        & \nodata & 10.72 & 4.75 & 3.50 & 1.65  & 8.41 & \nodata & SB(rs)bc \cr 
\hline 
NGC 1097 & 13.60 & 10.76 & 2.60 & 6.60 & 4.70 & 8.84 & YCl2.5rGnR & (R\am)SB(rs,bl,nr)ab pec G \cr 
NGC 2903 & 10.00 & 10.63 & 3.70 & 3.40 & 3.10 & 8.79 & YCl1.5rGnR & (R\am)SB(rs,nr)b M \cr 
NGC 5236 & 4.90 & 10.53 & 3.50 & 2.80 & 4.20 & 8.87 & YCl1.5rGnR & SAB(s,nr)c M \cr 
\enddata  
\tablecomments{ 1. L(CO) is the luminosity of the \jj10 line of CO, restricted to $\rgal/\reff\ \leq 1.52$. 2.  Type is the Stuber type, while CVRHS is the Comprehensive de Vaucouleurs revised Hubble-Sandage type from Table 6 of the \citet{2015ApJS..217...32B} analysis of Spitzer data. The letter following the CVRHS class indicates grand-design (G) or Multi-armed (M) from the additional notes column in Table 6.  } 
 \end{deluxetable*}

The properties of the eight galaxies are listed in Table \ref{tab:bigdips}.
Those of the shallow-dip galaxies are listed below the horizontal line.
The distance, stellar mass, and SFR are taken from Table A1 of 
\citetalias{2022AJ....164...43S}, 
while effective (half-mass) radius (\reff) is taken from
\citet{2021ApJS..257...43L}. 
The molecular morphological type and \rbar\ are from 
\citet{2023A&A...676A.113S},
Tables 1 and D.1, respectively.
The luminosity of the CO \jj10\ has been calculated from the radial profiles in the figures but only out to $\rgal/\reff = 1.52$ so that we are comparing only to the same effective radius.
For the \mw, the stellar mass (\mstar) and ``light-weighted" half-mass radius  are from 
\citet{2025ApJ...990..203I},
the SFR is from \citet{2015ApJ...806...96L}, the bar length is from \citet{2023MNRAS.520.4779L}, and the CVRHS classification is from 
\citet{2002SSRv..100..129G}.
The CO luminosity is again only that inside $\rgal/\reff = 1.52$ for fair comparison. As expected from Figure \ref{bigdipgals}, NGC 4548 has the CO luminosity most similar to that of the \mw, which has the lowest CO luminosity. 
\added{As discussed in \S \ref{sec:caveats}, the CO luminosity for the \mw\ may be underestimated because of beam dilution of the more distant parts of the Galaxy.
}

So, is the \mw\ odd? For the 26 galaxies with data out to $\rgal = 8$ kpc, 8 have class YClxxGnR; 5 of those show big dips (Figure \ref{bigdipgals}), two have small, \added{shallow} dips, and one has a large, shallow dip (Figure \ref{fig:smalldips}). 
Out of the entire sample considered by
\citet{2023A&A...676A.113S},
one-third have clear bars (C) and half of those have long bars (l); about one-quarter have grand-design spiral arms (G). The full classification of YCLxxGnR is shared by only 9 galaxies out of 79 in Table 1 of 
\citet{2023A&A...676A.113S}. Requiring only YCl adds 4 more galaxies, three classed as flocculent and one as multi-armed.
The molecular morphology of the \mw\ is rather unusual among PHANGS galaxies, but it is quite typical of galaxies with a particular molecular morphological classification that describes about 10\% of PHANGS galaxies.

\section{Discussion}\label{disc}

The distinctive big dip in the radial distribution of molecular gas and star formation in the \mw\ is seen in a small fraction of galaxies in the PHANGS study. All of those galaxies have a characteristic classification of their molecular morphology
\citep{2023A&A...676A.113S},
featuring clearly defined, long bars and grand-design spirals outside the bars. These data suggest that the \mw\ shares this classification. The presence of the bar agrees well with recent analysis of stars in the inner \mw, with evidence for a bar of length similar to that of the dip in molecular gas. The analysis of stars is complicated by the existence of different age and metallicity populations in the inner \mw, but analyses of the younger, more metal-rich stars find strong evidence for a bar
\citep{2021A&A...656A.156Q}.

Because all the galaxies with dips have bars with lengths that correlate with the extent of the dip, this analysis provides a new line of support for the hypothesis of a bar with radius about 4 kpc in the \mw, as suggested by 
\citet{1991ApJ...379..631B}
and many later papers
\citep[e.g.,][]{2023MNRAS.520.4779L}.
The fact that the similar galaxies also have grand-design spirals \added{is} less relevant \added{as discussed later} because the big dip is likely caused by clearance of molecular gas by the bar.

The galaxies in Table 1, especially those above the horizontal line, are candidates for \mw\ {\it molecular} analogues. In particular, the radial profile of \mean{\ico} for NGC 4548 is nearly identical to that of the \mw. These may provide alternatives to analogues identified based on stellar properties. None of the galaxies in Table 1 appear in a list of 176 analogues based on morphology of spiral arms, bar, and small bulge 
\citep{2019MNRAS.489.5030F}.
\added{To a large extent, the lack of overlap is caused by the northern hemisphere bias of the SDSS sample used by
\citet{2019MNRAS.489.5030F} 
versus the southern hemisphere bias of PHANGS (and any ALMA survey). Only three of our 8 galaxies are in the SDSS survey and one of those lies outside the stellar mass limits imposed by
\citet{2019MNRAS.489.5030F}.
}

\begin{figure*}[h!]
\includegraphics[width=0.94\textwidth]{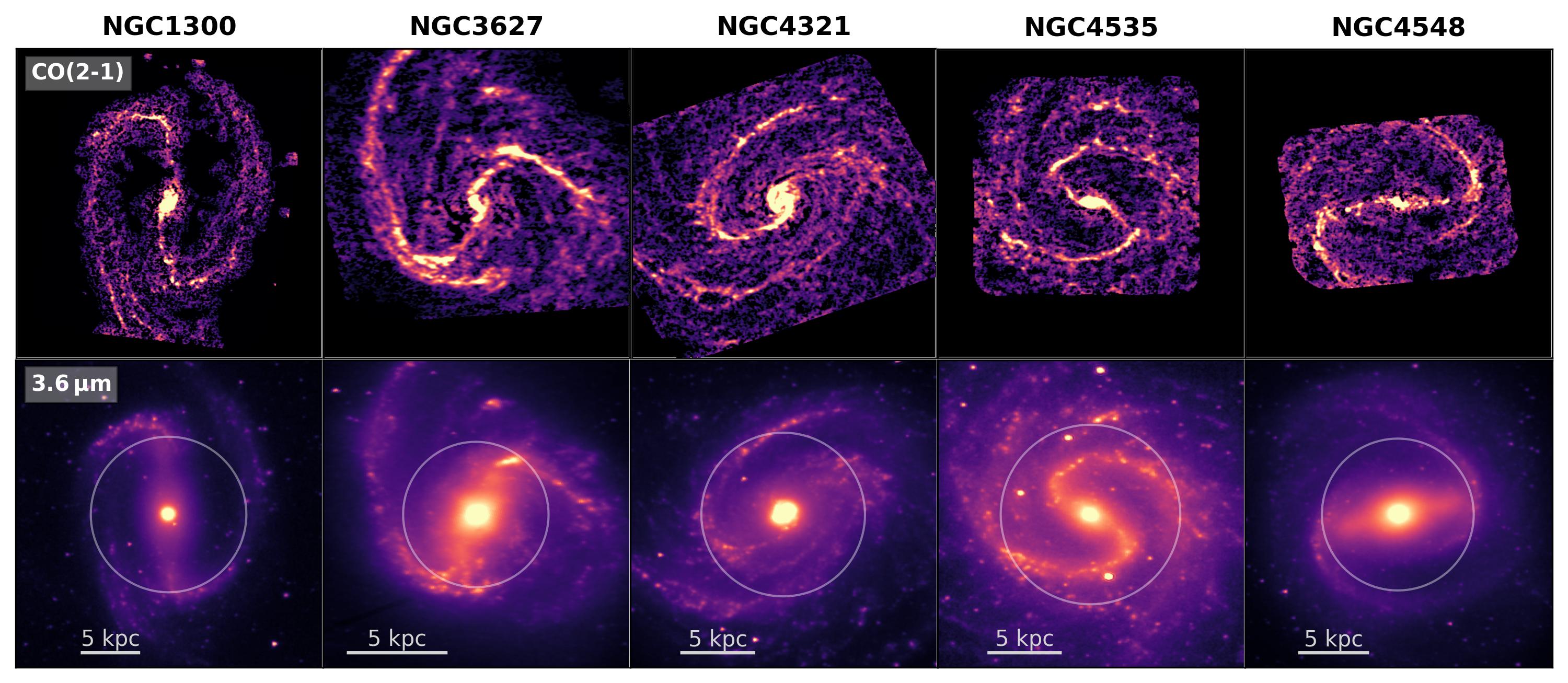}
\caption{Images of the external galaxies in Table 1 with deep dips in the radial profile of \mean{\ico}. 
The top images show the integrated intensity maps of CO \jj21 at their native resolution from the PHANGS-ALMA public data release. These so-called ``broad-mom-0'' maps are obtained from integrating along the spectral dimension and masking out noise as described in \cite{leroy_2021_pipeline}. 
The images are deprojected and derotated using inclination and position angles from \cite{2021ApJS..257...43L}.
The bottom images display the 3.6 \micron\ light as imaged by Spitzer as part of the Spitzer Survey of Stellar Structure in Galaxies \citep[S$^4$G,][]{Sheth_2010PASP..122.1397S} and are deprojected and derotated as well. 
We indicate the galaxy effective radius as a white circle ($R_\mathrm{e}$, Table~\ref{tab:bigdips}).
}
\label{fig:images}
\end{figure*}

\begin{figure*}[h!]
\includegraphics[width=0.56\textwidth]{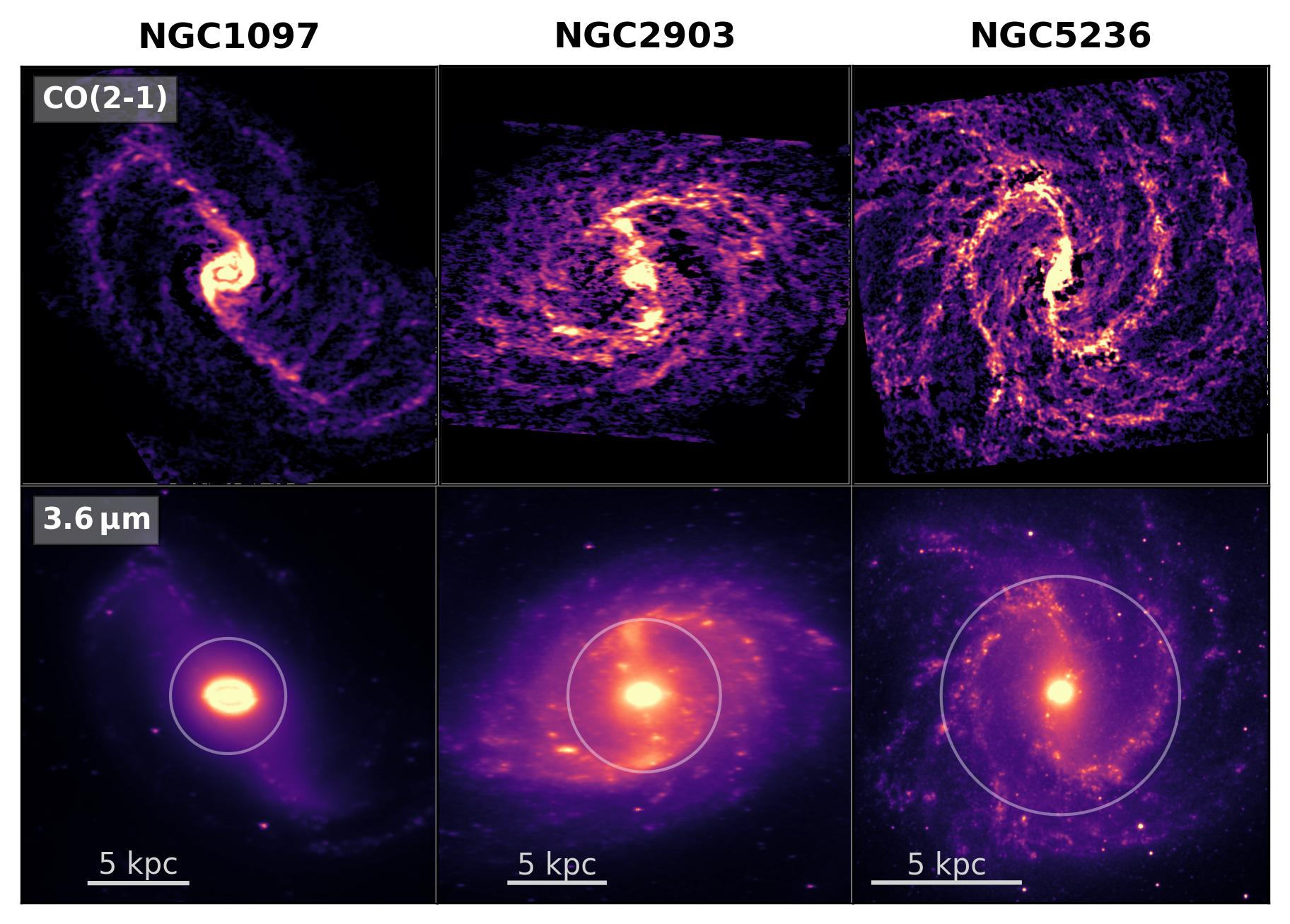}
\caption{Images of the external galaxies in Table 1 with more shallow dips. 
The top images are in CO \jj21, while the bottom images are in 3.6 \micron\ light as imaged by Spitzer.
Same as Figure~\ref{fig:images}.
}
\label{fig:images2}
\end{figure*}

\added{Images of the eight galaxies are shown in Figures \ref{fig:images} and \ref{fig:images2} with CO \jj21\ on top and Spitzer 3.6 \micron\ images below. The variety of spiral arm structures in the images and sometimes differing spiral arm classes between Stuber and CVRHS reinforces our warning not to over-interpret the fact that all the galaxies with big dips have a Stuber type that includes a grand-design spiral. The key similarity to the \mw\ is the bar clearing a gap, while the structure of the outer \mw\ is less relevant. Finding evidence for spiral arms in the \mw\ molecular gas is difficult, even with suggested arms to guide the eye \citep[see figures in][]{MD17}, and recent work favors a flocculent galaxy in atomic gas \citep{2025AJ....170..355B}.

The evidence for a flocculent galaxy extends beyond gas. Large astrometric surveys, like Gaia, have paved the way for detailed mapping of the stellar content of the Milky Way and current studies of a wide range of stellar tracers point towards a Milky Way that is flocculent rather than grand-design in structure \citep[e.g.,][]{Quillen2018, Hunt2025}. For example, the distribution of open clusters \citep[e.g.,][]{Castro-Ginard2021}, young, upper main sequence stars and masers \citep[e.g.,][]{Xu2023} and an older stellar population \citep[e.g.,][]{Quillen2018} disfavor a grand-design Milky Way. 
}

\begin{figure}
    \centering
    \includegraphics[width=0.4\textwidth]{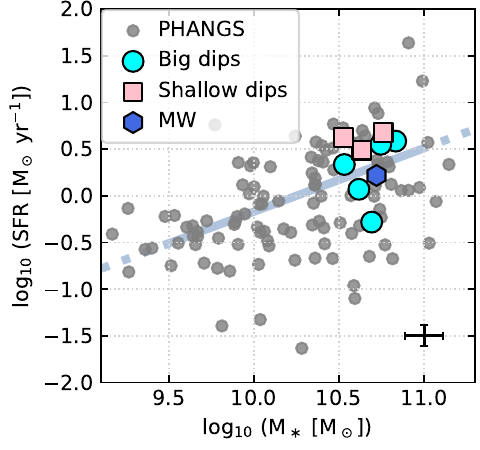}
    \caption{Top panel: Global SFR as function of stellar mass for the PHANGS-ALMA sample (grey points), the Milky Way (blue hexagon), the PHANGS galaxies with big dips (cyan circles), and those with shallow dips (pink squares). 
    The star-forming main sequence according to \citet{2019ApJS..244...24L} is added for reference (solid line, extrapolated to larger and smaller stellar masses with a dotted line).
    The values are listed in Table~\ref{tab:bigdips}.
    Error bars \citep[0.114 dex for all properties from][]{2021ApJS..257...43L} are shown in the bottom right corner. 
    }
    \label{fig:Mainsequence}
\end{figure}

How do overall properties of the molecular analogues and the \mw\ compare to the full PHANGS sample?
Figure~\ref{fig:Mainsequence} shows the global SFR as a function of stellar mass of the Milky Way and PHANGS galaxies, including the 8 galaxies of similar morphology to the Milky Way from Table~\ref{tab:bigdips}.
There is no significant ($\lesssim3\sigma$) difference between average SFR and average stellar mass of the  5 big-dip galaxies (Fig.~\ref{bigdipgals}) and the three shallow-dip galaxies (Fig.~\ref{fig:smalldips}).
The Milky Way global properties agree with both sub-samples. 

Despite being selected based only on their radial \mean{\ico} profiles, all those galaxies including the Milky Way are at 
the higher stellar mass end of the PHANGS sample (Figure~\ref{fig:Mainsequence}). 
In particular, they lie above the stellar mass of 1\ee{10} \msun, identified as important for the morphology of the gas inside the bar 
\citep{2025A&A...700A...3V}.

Comparison to recent studies of the \mw\ in stars may also be of interest. 
Recent analyses differ, but we will compare to the detailed analysis by
\citet{2025ApJ...990..203I}.
They fit the total stellar surface density to an exponential with scale-length, $\rdisk = 2.83 \pm 0.2$ kpc. This is somewhat longer than the molecular data, fit to 5-14 kpc, of $\rdisk = 2.26 \pm 0.14$ found by \citetalias{2025ApJ...980..216E}.
\citet{2025ApJ...990..203I} 
also comment on a possible truncation around 13 kpc that they suggest may be linked to the gas density dropping below a critical threshold for star formation. However, \citetalias{2025ApJ...980..216E} see no truncation beyond 13 kpc in either the molecular or star formation surface densities, within the limits imposed by stochastic variations.

While few PHANGS galaxies have data much beyond 2 \reff, the radial profiles in Figures \ref{bigdipgals} and \ref{fig:smalldips} suggest that \mean{\ico} declines more slowly in those galaxies than in the \mw.
Several factors may be at work. Lower sensitivity to small clouds on the far side of the \mw\ could cause an underestimate of $\mean{\ico}$, but this would be a factor of two at most.

\added{
\section{Caveats}\label{sec:caveats}

The existence of the big dip in the \mw\ can be questioned on two grounds: firstly, clouds in that region may be incorrectly placed elsewhere by the kinematic distance method; secondly, beam dilution might cause underestimation of the mean CO intensity at the distance of the dip. 

Concern about the misplacement of clouds in fact first motivated us to look at the PHANGS sample. If no other galaxies had such a dip, we would have been very suspicious about its occurence in the \mw. The fact that PHANGS galaxies with bars similar to the bar suggested for the \mw\ have dips is the best argument that the dip is real. However, it could still be less deep than shown in Figure \ref{bigdipgals} if many clouds are misplaced. In particular, CO gas flowing along the bar has been identified by
\citet{2008A&A...477L..21M}
using extinction and by
\citet{2019MNRAS.488.4663S} using CO emission with distinctive locations in the $(l,v)$ and $(l,b)$ planes. We identified all clouds in those ranges and checked their \rgal. Most are in fact placed in the region of the dip, but a few are placed in the CMZ. Placing all of them in the dip region has only a modest effect on the dip. More extreme experiments of moving clouds with uncertain distances into the dip region can make the dip more shallow but do not eliminate it.

Distance biases in the catalog were discussed in Appendix C of \citet{MD17}. Figure 24 of that paper shows the limits on cloud radius and mass as a function of distance. For a distance of 4-8 kpc (the front side of the bar region), the mass limit increases from 1\ee3 to 5\ee3 \msun. Most clouds lie well above this limit, though there is a group that cluster along the limit. Because most of the mass is in more massive clouds, missing smaller clouds probably has only a small effect on the depth of the dip.

Improved information about the dip will come from better resolution CO surveys underway and improved models of the kinematics of the bar region.
}

\section{Summary}\label{summary}

The key results are the following.

\begin{itemize}

\item The \mw\ has a big dip in the radial profile of \mean{\ico}.
\item This dip is also seen in a subsample of PHANGS galaxies, all with a characteristic Stuber classification of YCLxxGnR, meaning clearly defined, long bars and grand-design spirals outside the bar.
\item This similarity suggests that an external observer would assign a Stuber class of YCLxx?nR to the \mw, 
\added{with the question mark reflecting the current state of uncertainty about the appearance of the galaxy outside the bar.
}
\item The galaxies with this classification provide a set of \mw\ analogues selected in a very different way from other analogues.
\item The bar length correlates well with the extent of the dip for these galaxies.
\item The Milky Way matches well in \mstar\ and SFR with those galaxies despite their selection being based only on the morphology.

\item All morphology-matched galaxies are at the higher stellar mass end of PHANGS sample. 
\end{itemize}

\begin{acknowledgments}

\added{We thank the referee for a prompt, yet insightful and constructive report.
}
This research used the Canadian Advanced Network For Astronomy Research (CANFAR) operated in partnership by the Canadian Astronomy Data Centre and The Digital Research Alliance of Canada with support from the National Research Council of Canada the Canadian Space Agency, CANARIE and the Canadian Foundation for Innovation. NJE is grateful to the Astronomy Department of the University of
Texas for research support, INAF-IAPS (Roma, Italy) for
hospitality during a visit, to H. Khan-Farooki, along with two
anonymous cornea donors, for the gift of sight.
DE acknowledges funding from INAF Mini Grant 2022 “The Multi-Fractal Structure of the InterStellar Medium (MSISM).”
S.K.S is supported by an International Research Fellowship of the Japan Society for the Promotion of Science (JSPS).
KH is partially supported by NSF AST-2407975. KH acknowledges support from the Wootton Center for Astrophysical Plasma Properties, a U.S. Department of Energy NNSA Stewardship Science Academic Alliance Center of Excellence supported under award numbers DE-NA0003843 and DE-NA0004149, from the United States Department of Energy under grant DE-SC0010623.

\end{acknowledgments}



\bibliography{cite}{}
\bibliographystyle{aasjournal}

\end{document}